\documentclass[doublecol]{epl2} 

\usepackage[colorlinks=true,linkcolor=blue,anchorcolor=blue,citecolor=blue,urlcolor=blue]{hyperref}

\title{Kibble-Zurek dynamics in an array of coupled binary Bose condensates}

\author{Jun Xu\inst{1,2,3}, Shuyuan Wu\inst{1,2}, Xizhou Qin\inst{1}, Jiahao Huang\inst{1,2}, Yongguan Ke\inst{1,2}, Honghua Zhong\inst{1,4} \and Chaohong Lee\inst{1,2}\thanks{E-mail: \email{chleecn@gmail.com}}}
\shortauthor{J. Xu \etal}

\institute{
  \inst{1} School of Physics and Astronomy, Sun Yat-Sen University, Guangzhou 510275, China\\
  \inst{2} State Key Laboratory of Optoelectronic Materials and Technologies, Sun Yat-Sen University, Guangzhou 510275, China\\
  \inst{3} Center of Experimental Teaching for Common Basic Courses, South China Agriculture University, Guangzhou 510642,China\\
  \inst{4}Department of Physics, Jishou University, Jishou 416000, China
}


\abstract{Universal dynamics of spontaneous symmetry breaking is central to understanding the universal behavior of spontaneous defect formation in various system from the early universe, condensed-matter systems to ultracold atomic systems.
We explore the universal real-time dynamics in an array of coupled binary atomic Bose-Einstein condensates in optical lattices, which undergo a spontaneous symmetry breaking from the symmetric Rabi oscillation to the broken-symmetry self-trapping.
In addition to Goldstone modes, there exist gapped Higgs mode whose excitation gap vanishes at the critical point.
In the slow passage through the critical point, we analytically find that the symmetry-breaking dynamics obeys the Kibble-Zurek mechanism.
From the scalings of bifurcation delay and domain formation, we numerically extract two Kibble-Zurek exponents $b_{1}=\nu/(1+\nu z)$ and $b_{2}=1/(1+\nu z)$, which give the static correlation-length critical exponent $\nu$ and the dynamic critical exponent $z$.
Our approach provides an efficient way to simultaneous determination of the critical exponents $\nu$ and $z$ for a continuous phase transition.}
\begin{document}
\maketitle
\textbf{Introduction.}\ -\ Symmetry, one of the most important concepts in physical sciences, describes the invariance of geometric configurations or intrinsic laws under specified transformations. Spontaneous symmetry breaking (SSB)~\cite{SSB} plays a significant role in many fundamental phenomena. Through a SSB, the state of a system evolves from symmetric to asymmetric, even though the underlying dynamic equations are still invariant under the symmetry transformation. It has been demonstrated that the critical dynamics of continuous SSB obeys the Kibble-Zurek (KZ) mechanism~\cite{Bunkov2000,Zurek1996,Kibble2007, Dziarmaga2010, Polkovnikov2011,Campo}, in which defect excitations are spontaneously generated. Universal critical dynamics obeying KZ mechanism have been found in various systems, such as, the early universe~\cite{Kibble}, superfluid helium~\cite{Zurek}, electronic unstable crystal~\cite{Yusupov}, multiferroic hexagonal maganites~\cite{Lin}, nematic liquid crystal~\cite{Nikkhou}, ion crystal~\cite{Ulm, Pyka}, and atomic Bose-Einstein condensates (BECs)~\cite{Damski2007, Weiler2008, Lee2009, Zurek2009, Damski2010, Witkowska, Sabbatini2011,Sabbatini2012, Matuszewski,Anquez,Dziarmaga2008, Chen2011, Su2013, Lamporesi2013, Hofmann2014, Navon2015}.

Attribute to their high controllability and robust quantum coherence, beyond testing KZ mechanism in thermodynamic phase transitions~\cite{Weiler2008, Zurek2009, Damski2010, Witkowska, Su2013, Lamporesi2013, Navon2015}, atomic BECs provide new opportunities for exploring KZ mechanism in quantum phase transitions~\cite{Damski2007, Lee2009, Sabbatini2011,Sabbatini2012,Matuszewski,Anquez, Dziarmaga2008, Chen2011, Hofmann2014}. To extract universal scalings, several local or global quantities are analyzed. The most used quantity is the total number of generated defects, such as, solitons~\cite{Zurek2009, Damski2010, Witkowska, Lamporesi2013}, vortices~\cite{Weiler2008} and domains~\cite{Damski2007, Sabbatini2011,Sabbatini2012,Matuszewski, Hofmann2014, Navon2015}. The total number of generated defects is a global quantity dependent on the whole system. Usually, the scaling of a global quantity may only give one KZ exponent, which is a combination of the static correlation-length critical exponent $\nu$ and the dynamic critical exponent $z$. There are also some works analyzing local quantity such as magnetization~\cite{Damski2008} and spin polarization~\cite{Lee2009}, which give another KZ exponent (also a combination of the critical exponents $\nu$ and $z$). Therefore, the critical exponents $\nu$ and $z$ can be simultaneously determined by analyzing both local and global quantities. The determination of the critical exponents $\nu$ and $z$ is very important to characterize the universality of phase transitions. Up to now, there is still no report on the simultaneous determination of the critical exponents $\nu$ and $z$ via analyzing real-time critical dynamics.

In this Letter, we investigate the real-time critical dynamics in an array of coupled binary BECs in optical lattices, which undergo a dynamical SSB from the symmetric Rabi oscillation to the broken-symmetry self-trapping. By analyzing the instantaneous Bogoliubov excitation gaps, we analytically find that the critical dynamics obeys the KZ mechanism. To extract universal scaling exponents, we simultaneously analyze a global quantity, the total number of generated domains, and a local quantity, the bifurcation delay of spin polarization. The scaling for the domain formation gives one KZ exponent, $b_{1}=\nu/(1+\nu z)$, which is a combination of the critical exponents $\nu$ and $z$. The scaling for the bifurcation delay gives another KZ exponent, $b_{2}=1/(1+\nu z)$, which is also a combination of the critical exponents $\nu$ and $z$. At the first time, from the two KZ exponents $(b_{1}, b_{2})$, we simultaneously determine the critical exponents $\nu=1/2$ and $z=1$. Through analyzing real-time critical dynamics, our approach can be used to simultaneously determine the critical exponents $\nu$ and $z$ for all continuous phase transitions.

\textbf{Model.}\ -\ We consider an array of coupled binary atomic BECs trapped in a deep one-dimensional state-independent optical lattice potential, see Fig.~1.This system can be realized by loading a two-component $^{87}$Rb BEC into a one-dimensional optical lattice and the two atomic hyperfine states $|1\rangle = \left|F=1,m_{f}=-1\right\rangle$ and $|2\rangle = \left|F=2,m_{f}=1\right\rangle$ are coupled by Raman lasers or radio-frequency fields~\cite{Hall}.In the mean-field theory, the system can be described the Hamiltonian (in the units of the Planck constant $\hbar=1$),
\begin{eqnarray}
H=\!\!\!\!\!\!\!\!\!&&-\frac{\Omega(t)}{2}\sum_{l}\left(\psi_{l,1}^*\psi_{l,2}+\psi_{l,2}^*\psi_{l,1}\right)\nonumber\\ &&+\frac{\delta}{2}\sum_{l}\left(\psi_{l,2}^*\psi_{l,2}-\psi_{l,1}^*\psi_{l,1}\right)\nonumber\\
&&-J\sum_{l}\sum_{\sigma=1}^2\left(\psi_{l,\sigma}^*\psi_{l+1,\sigma}+\psi_{l+1,\sigma}^*\psi_{l,\sigma}\right)\nonumber\\
&&+\sum_{l}\sum_{\sigma=1}^2\frac{g_{\sigma\sigma}}{2}|\psi_{l,\sigma}|^{4}+g_{12}\sum_{l}|\psi_{l,1}|^{2}|\psi_{l,2}|^{2}.
\end{eqnarray}

with the complex numbers $\psi_{l,\sigma}$ denoting the order parameters for the $\sigma$-component in the $l$-th lattice site.
Here, $J \geq 0$ describe the inter-lattice hopping strength, $\Omega  \geq 0$ stands for the coupling strength, $\delta  \geq 0$ labels the detuning, and $g_{\sigma\sigma^{\prime}}$ denote the atom-atom interaction.

\begin{figure}[h]
\includegraphics[width=1.0\columnwidth]{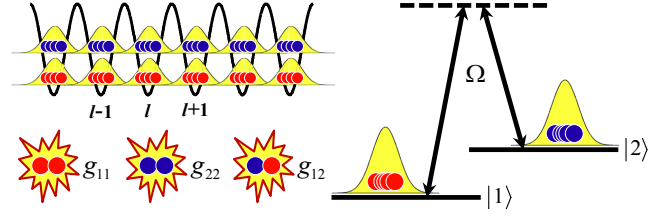}
\caption{Schematic diagram. An array of coupled binary atomic BECs are confined in a one-dimensional optical lattice potential. The two hyperfine levels are coupled by Raman lasers. The atom-atom interaction are dominated by s-wave scattering.}
\label{fig:SchematicDiagram}
\end{figure}

\textbf{Static Spontaneous Symmetry Breaking.}\ -\ The time-evolution obeys the discrete Gross-Pitaevskii equation (DGPE), $i\frac{d\psi_{l,\sigma}}{dt}=\frac{\partial H}{\partial \psi_{l,\sigma}^{*}}$, in which $\psi_{l,\sigma}^{*}$ being the complex conjugates of $\psi_{l,\sigma}$.
Under periodic boundary condition ($\psi_{L+1,\sigma}=\psi_{1,\sigma}$), given the lattice number $L$ and the particle number $N$, it is easy to find that the system has stationary states $\psi_{l,\sigma}=C_{\sigma}\textup{e}^{-i(\mu t-\phi_{\sigma})}$ with the chemical potential $\mu$, the lattice indices $l=\{1,2, \cdots, L-1, L\}$, and the amplitudes $C_{\sigma}$ satisfying the normalization condition $|C_1|^2+|C_2|^2=N/L$.
If the collision shift is balanced by the detuning, i.e. $\delta=(g_{11}-g_{22})N/(2L)$, the Hamiltonian (1) is invariant under the transformation $\psi_{l,1} \leftrightarrow \psi_{l,2}$.
Dependent on the ratio $\left|\frac{N\Delta}{2L\Omega}\right|$ and the sign of $\Delta=g_{11}+g_{22}-2g_{12}$, the two amplitudes are equal or unequal.
If $\Delta<0$, varying $\left|\frac{N\Delta}{2L\Omega}\right|<1$ to $\left|\frac{N\Delta}{2L\Omega}\right|>1$, the in-phase solutions ($\phi_1=\phi_2$) change from the symmetric states with
$C_1=C_2=\sqrt{N/(2L)}$
to the broken-symmetry states with
$C_1=\left[\frac{N}{2L}\left(1\pm\sqrt{1-\frac{4\Omega^2L^2}{\Delta^2N^2}}\right)\right]^{\frac{1}{2}}$ and
$C_2=\left[\frac{N}{2L}\left(1\mp\sqrt{1-\frac{4\Omega^2L^2}{\Delta^2N^2}}\right)\right]^{\frac{1}{2}}$.
If $\Delta>0$, the anti-phase states ($\phi_1-\phi_2=\pi$) show similar SSB.
In the limit of only one lattice site, the above SSB becomes the transition from Rabi oscillation to macroscopic quantum self-trapping~\cite{Lee2009,Lee2004,Lee2012}.
Mathematically, such a SSB corresponds to a pitchfork bifurcation from single- to bi-stability.
Without loss of generality, below we will only discuss the SSB of in-phase states.

\textbf{Dynamical Spontaneous Symmetry Breaking.}\ -\ Now we analyze the dynamical SSB across the critical point $\left|\frac{N\Delta}{2L\Omega}\right|=1$.
This dynamical process sensitively depends on the instantaneous excitation gap $\omega$, which determines two characteristic time scales~\cite{Damski2007,Lee2009}: the reaction time $\tau_{r} \sim 1/\omega$ and the transition time $\tau_{t} \sim \omega/\left|d\omega/dt\right|$.
The reaction time $\tau_{r}$ and the transition time $\tau_{t}$ describe how fast the system follows its instantaneous eigenstates and how fast the system is driven, respectively.
The system undergoes adiabatic evolution if $\tau_{r} < \tau_{t}$. Otherwise, if the system has no sufficient time to follow the instantaneous eigenstates, non-adiabatic excitations appear.

We employ the Bogoliubov theory~\cite{RevModPhys.77.187} to obtain the instantaneous excitation gap. The perturbed stationary states read as
\begin{equation}
\psi_{l,\sigma}(t)=\left[C_{\sigma}+\delta\psi_{l,\sigma}(t)\right]\textup{e}^{-i\mu t},
\end{equation}
in which the perturbation terms are written as
\begin{equation}
\delta\psi_{l,\sigma}(t)=\sum_q\left[u_{q,\sigma}\textup{e}^{+i(ql-\omega_q t)}+v_{q,\sigma}^*\textup{e}^{-i(ql-\omega_q t)}\right].
\end{equation}
Here $\omega_q$ denotes the excitation gap and the quasimomenta are given as $q=2\pi\left(\frac{n}{L}-\frac{1}{2}\right)$ with the integers $n=\left(0, 1, \cdots, L-2, L-1\right)$.
Inserting the perturbed stationary states into the DGPE and comparing the coefficients for terms of $\textup{e}^{+i(ql-\omega_q t)}$ and $\textup{e}^{-i(ql-\omega_q t)}$, we obtain the Bogoliubov-de Gennes equations for the complex amplitudes $\left(u_{q,\sigma}, v_{q,\sigma}\right)$.
The excitation gaps $\omega_q$ can then be obtained by diagonalizing the Bogoliubov-de Gennes equations.

We find that there are two typical kinds of excitations: Goldstone modes and Higgs modes~\cite{Anderson}.
The Goldstone modes originate from fluctuations in the density sector and become gapless in the long-wavelength limit ($q=0$).
The Higgs modes originate from fluctuations in the spin sector and have nonzero energy gaps exception for the critical point $\left|\frac{N\Delta}{2L\Omega}\right|=1$.
In the long-wavelength limit ($q=0$), as the Goldstone modes are trivial gapless excitations ($\omega_0^G=0$), the nontrivial excitations are the Higgs modes whose excitation gaps $\omega_0^H=\sqrt{\Omega(\Omega-\Omega_c)}$ if $\Omega \geq \Omega_c$ or $\omega_0^H=\sqrt{\Omega_c^2-\Omega^2}$ if $\Omega \leq \Omega_c$.
The excitation gap $\omega_0^H$ of the Higgs modes gradually vanishes when $\Omega$ approaches to its critical value $\Omega_c=\left|\frac{N\Delta}{2L}\right|$.
Due to the gapless Higgs mode at the critical point, non-adiabatic excitations always appear when the system is quenched across the critical point.
In Fig.~2, we show the excitation spectrum around the critical point.

\begin{figure}[h]
\includegraphics[width=1.0\columnwidth]{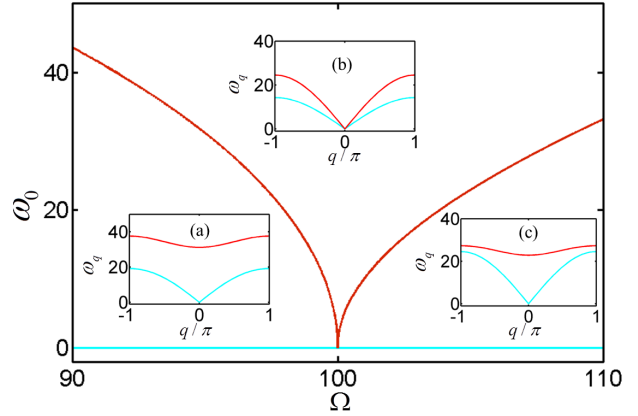}
\caption{Excitation spectrum around the critical point. The parameters are chosen as $N=100000$, $L =100$, $J=0.5$, $g_{11}=g_{22}=1$, $g_{12}=2$ and the quasimomenta $q=0$. The critical point for these parameters is $\Omega=\Omega_c=100$. Insets: The excitation spectrum versus the quasimomenta $q$ for (a) $\Omega=95$, (b) $\Omega=100$ and (c) $\Omega=105$. Here the red and blue curves correspond to Higgs and Goldstone modes, respectively.}
\label{fig:BdG}
\end{figure}

Given the instantaneous excitation gaps, we may derive the universal dynamic mechanism for the process across the critical point.
Near the critical point, given the dimensionless distance to the critical point $\varepsilon (t)$, the energy gap vanishes as $\omega (t) \sim \left|\varepsilon (t)\right|^{z\nu}$, the correlation length diverges as $\xi(t)\sim \left|\varepsilon (t)\right|^{-\nu}$, the reaction time $\tau_{r}(t) \sim \omega^{-1} \sim |\varepsilon(t)|^{-z\nu}$ and the transition time $\tau_t (t) \sim \omega (t)/\left|d\omega(t)/dt\right| \sim \varepsilon (t)/\left|d\varepsilon(t)/dt\right|$.
The adiabaticity breaks down near the freeze-out time $t=\hat{t}$ when the transition rate $\left|d\varepsilon/dt\right|/\varepsilon$ equals the energy gap $\omega$, that is, $\tau_r (\hat{t}) =\tau_t(\hat{t})$.
In our system, we assume the coupling strength $\Omega$ is linearly quenched, that is, $\Omega=\Omega_c(1 \pm t/\tau_Q)$ and $\varepsilon=\left|(\Omega(t)-\Omega_{c})/\Omega_{c}\right| = |t|/\tau_{Q}$ with $\tau_Q$ being the quenching time.
Given $\omega=\omega_0^H$ for nontrivial excitations, in the vicinity of $\Omega_c$, we have the reaction time
\begin{equation}
\tau_r \sim \frac{1}{\omega} \sim
\left\{
\begin{array}{ll}
\left[\Omega_c \sqrt{\varepsilon(1+\varepsilon)}\right]^{-1} \sim \varepsilon^{-\frac{1}{2}}, \textup{~if~} \Omega \geq \Omega_c, \\
\left[\Omega_c \sqrt{\varepsilon(2-\varepsilon)}\right]^{-1} \sim \varepsilon^{-\frac{1}{2}}, \textup{~if~} \Omega \leq \Omega_c.
\end{array}
\right.
\end{equation}
and the transition time
\begin{equation}
\tau_{t} \sim \frac{\omega}{\left|d\omega/dt\right|} \sim \frac{\varepsilon}{\left|d\varepsilon/dt\right|} \sim t.
\end{equation}
Obviously, comparing Eq.~(4) with $\tau_{r} \sim |\varepsilon |^{-z\nu}$, we have $z\nu=1/2$.
At the freeze-out time $t=\hat{t}$, from $\tau_r (\hat{t}) = \tau_t(\hat{t})$, we find the following universal scalings,
\begin{equation}
\left|\hat{t}\right| \sim \tau_Q^{1/3},~~~\hat{\varepsilon}=\varepsilon (\hat{t}) \sim \tau_Q^{-2/3},
\end{equation}
with respect to the quenching time $\tau_Q$.
Comparing these scalings with the KZ scalings~\cite{Bunkov2000, Dziarmaga2010, Polkovnikov2011,Campo}: $\left|\hat{t}\right| \sim \tau_Q^{z\nu/(1+z\nu)}$ and $\hat{\varepsilon} \sim \tau_Q^{-1/(1+z\nu)}$, we obtain $z\nu=1/2$.
According to $\xi(t)\sim \left|\varepsilon (t)\right|^{-\nu}$, the correlation length at the freeze-out time $t=\hat{t}$ scales as $\hat{\xi} = \xi(\hat{t}) \sim \left|\hat{\varepsilon}\right|^{-\nu} \sim \tau_Q^{\nu/(1+z\nu)}$ and the defect number scales as $N_d \sim \hat{\xi}^{-1} \sim \tau_Q^{-\nu/(1+z\nu)}$.

\textbf{Numerical Scalings.}\ -\ To examine the above analytical KZ scalings, we numerically simulate the slow passage through the critical point and extract two KZ exponents.
Most of previous works extract only one KZ exponent via analyzing either global~\cite{Damski2007, Weiler2008, Zurek2009, Damski2010, Sabbatini2011,Sabbatini2012,Matuszewski,Lamporesi2013, Hofmann2014, Navon2015} or local~\cite{Damski2008, Lee2009} quantities of the critical dynamics.
Here, we extract two different KZ exponents from the scalings of the domain number (a global quantity) and the bifurcation delay (a local quantity) with respect to the quenching time $\tau_{Q}$.
In our calculation, given a set of parameters and an initial symmetric state, for each $\tau_Q$, we simulate the same quenching process [$\Omega=\Omega_c(1 - t/\tau_Q)$] for 200 different random perturbations.

In a specific run, from all local spin polarizations
$\rho_{l}^{(r)}=\left[n_1^{(r)}-n_2^{(r)}\right] /\left[n_1^{(r)}+n_2^{(r)}\right]$ (where $n_j^{(r)}=\left|\psi_{l,j}^{(r)}\right|^{2}$), the domain number $N_{d}^{(r)}$ is given by the number of zero points along the lattice direction.
Through analyzing the scaling of the average domain number,
\begin{equation}
N_{d}=\frac{1}{200}\sum_{r=1}^{200} N_{d}^{(r)} \sim \tau_Q^{-b_1} \sim \tau_Q^{-\nu/(1+z\nu)},
\end{equation}
with respect to $\tau_{Q}$, {we obtain the KZ exponent $b_1=\nu/(1+z\nu)=0.340 \pm 0.035$ from our numerical data for slow quenches ($10^2<\tau_{Q}<10^3$), see Fig.~3.
However, for rapid quenches ($\tau_{Q}<10^2$), the mean domain numbers deviate from the KZM prediction.
This is because that, for rapid quenches, the mean domain number increases significantly and the domain number may still decrease at the end of our integration time, and therefore the numerical data overestimates the domain number.
Similar overestimation appears in one-dimensional continuous two-component BEC systems~\cite{Sabbatini2011,Sabbatini2012}.}

\begin{figure}[h]
\includegraphics[width=1.0\columnwidth]{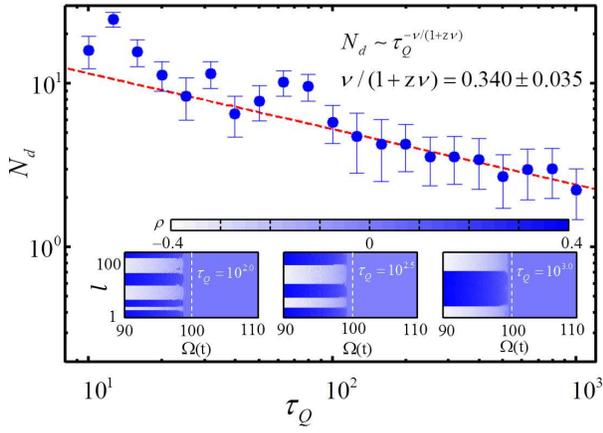}
\caption{Scaling of the domain number $N_d$ with respect to the quenching time $\tau_Q$. Insets: the time evolution of local spin polarizations $\rho$ for three different quenching time $\tau_Q$. The parameters are chosen as same as the ones for Fig.~2.}
\label{fig:Domain_scaling}
\end{figure}

Unlike a static bifurcation, in which the bifurcation occurs at the critical point, the dynamical bifurcation may take place after the system passes the critical point.
In a specific run, we determine the bifurcation delay,
\begin{equation}
b_{d}^{(r)}=\left|\Omega^{*}-\Omega_{c}\right| \sim \Omega_{c} \hat{\varepsilon},
\end{equation}
via analyzing spatial fluctuation of the local spin polarizations $\Delta _{\rho}^{(r)}=\left[\frac{1}{L}\sum_{l=1}^{L}\left(\rho_{l}^{(r)}\right)^{2} -\left(\frac{1}{L}\sum_{l=1}^{L}\rho_{l}^{(r)}\right)^{2}\right]^{\frac{1}{2}}$.
Before bifurcation, there is no spatial fluctuation, i.e. $\Delta_{\rho}^{(r)}=0$.
{The dynamical bifurcation appears at $\Omega^{*}$ where $\Delta_{\rho}^{(r)}$ reaches a small nonzero value $\delta_{\rho}$, which is chosen as $6\%$ in our calculation.
Based upon our calculations for different values of $\delta_{\rho}$, although the amplitude of $\Omega^{*}$ depends on the choice of $\delta_{\rho}$, the KZ exponent does not change.
Similar behavior has been found in the appearance of domain seeds in spinor BECs~\cite{Matuszewski}.}
Through analyzing the scaling of the average bifurcation delay,
\begin{equation}
b_{d}=\frac{1}{200}\sum_{r=1}^{200} b_{d}^{(r)} \sim \tau_Q^{-b_2} \sim \tau_Q^{-1/(1+z\nu)},
\end{equation}
with respect to $\tau_{Q}$, we obtain another KZ exponent $b_2=1/(1+z\nu)=0.670 \pm 0.004$, see Fig.~4.

\begin{figure}[h]
\includegraphics[width=1.0\columnwidth]{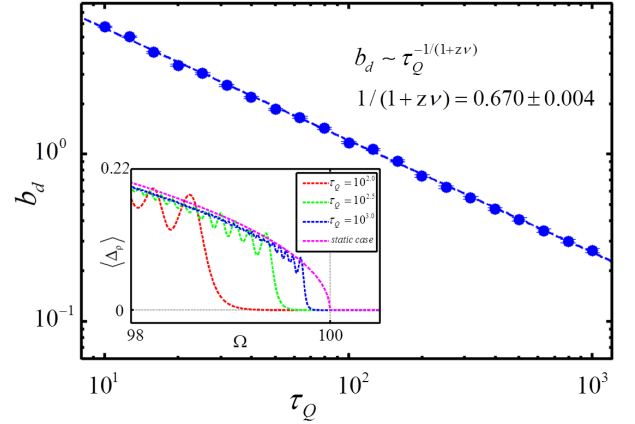}
\caption{Scaling of the bifurcation delay $b_d$ with respect to the quenching time $\tau_{Q}$. Inset: the averaged spatial fluctuations $\langle \Delta_{\rho}  \rangle= \left[\sum_{r=1}^{200}\Delta _{\rho}^{(r)}\right]/200$. The parameters are chosen as same as the ones for Fig.~2.}
\label{fig:Delay_scaling}
\end{figure}

With the two KZ exponents $b_1=0.340 \pm 0.035$ and $b_2=0.670 \pm 0.004$ extracted from the real-time critical dynamics, {we simultaneously derive $\nu=0.507 \pm 0.052$ and $z=0.971 \pm 0.101$ from the relations $b_1=\nu/(1+z\nu)$ and $b_2=1/(1+z\nu)$.}
These critical exponents and KZ exponents given by numerical simulation are well consistent with $\nu=1/2$ and $z=1$ and the analytical KZ scalings~(6) derived from the competition between the reaction time $\tau_{r}$ and the transition time $\tau_{t}$.
Furthermore, the analytical KZ scalings~(6) can only determine the value of $z\nu$, the above two KZ exponents can simultaneously determine the values of both $z$ and $\nu$.
Here the critical exponents $\nu$ and $z$ are obtained in the mean-field framework.
{Beyond mean-field framework, it has been experimentally reported that the thermodynamic Bose-Einstein condensation phase transition in a three-dimensional system belongs to the universality class of $\nu=2/3$ and $z=3/2$~\cite{Navon2015}.
In the mean-field framework, it has been experimentally found the one-dimensional binary atomic BEC supporting a quantum phase transition of $\nu=1/2$ and $z =1$~\cite{Nicklas2015}. Theoretically, both mean-field scalings and beyond-mean-field ones are discussed~\cite{Zurek1996,Zurek2009}.}

\textbf{Experimental Possibility.}\ -\ Based on the well-developed techniques of atomic BECs, it is possible to observe the above critical dynamics and determine the critical exponents.
In recent years, it has been experimentally demonstrated that the critical dynamics in a homogeneous Bose atomic gas undergoing a thermodynamic phase transition obeys the KZ mechanism~\cite{Navon2015}.
More recently, the experimental observation of scaling in the time evolution of a binary atomic BEC following a sudden quench into the vicinity of a quantum critical point has been reported~\cite{Nicklas2015}.
Based the experimental techniques used in~\cite{Hall, Nicklas2015,Kempen,Widera,Erhard}, an array of coupled binary BECs can be prepared by loading a binary $^{87}$Rb BEC into a one-dimensional optical lattice and the coupling between two involved hyperfine levels is coupled via Raman lasers or radio-frequency fields~\cite{Hall}.
{In such a system, one may tune both inter- and intra-component interactions via the techniques of Feshbach resonance~\cite{Kempen,Widera,Erhard}.}
The dynamical SSB from the symmetric Rabi oscillation to the broken-symmetry self-trapping is driven by tuning the intensity of the coupling fields.
The scalings of domain formation and bifurcation delay can be obtained by observing the local spin polarizations via absorption imaging.

\textbf{Conclusion.}\ -\ In summary, we have studied the real-time critical dynamics in an array of coupled binary BECs quenched through a continuous SSB.
The analytical KZ scalings are derived from the competition between the reaction time and the transition time.
We then numerically extract two KZ exponents from the scalings of domain formation and bifurcation delay with respect to the quenching time.
Our numerical scalings agree with the analytical KZ scalings.
{Based upon the relations $b_1=\nu/(1+z\nu)$ and $b_2=1/(1+z\nu)$, we simultaneously determine the two critical exponents $\left(\nu=0.507 \pm 0.052, z=0.971 \pm 0.101 \right)$ from our numerical estimations for the two KZ exponents $\left(b_1=0.340 \pm 0.035, b_2=0.670 \pm 0.004\right)$.}
Our approach for simultaneously determining the two critical exponents $\left(\nu, z\right)$ from real-time critical dynamics can be applied to other continuous phase transitions.
Based upon the present techniques of manipulating and probing binary atomic BECs~\cite{Hall, Nicklas2015}, the above real-time critical dynamics can be examined in experiments.

\acknowledgments
The authors acknowledges Q. Xie and H. Deng for discussions. This work is supported by the National Basic Research Program of China (NBRPC) under Grant No. 2012CB821305, the National Natural Science Foundation of China (NNSFC) under Grants No. 11374375 and 11574405, and the PhD Programs Foundation of Ministry of Education of China under Grant No. 20120171110022.

\end{document}